\documentclass[journal=ancac3,manuscript=article,layout=onecolumn]{achemso}
\usepackage{textcomp,gensymb,amsmath,xcolor,multirow,natbib,lineno,pdfpages}
\newcommand*\2{$_2$}
\newcommand{\txr}{\textcolor{black}}

\author{Nithin Abraham}
\affiliation[IISc]
{Department of Electrical Communication Engineering, Indian Institute of Science, Bangalore 560012, India}
\author{Kenji Watanabe}
\affiliation[NIMS1]
{Research Center for Functional Materials, National Institute for Materials Science, 1-1 Namiki, Tsukuba 305-0044, Japan}
\author{Takashi Taniguchi}
\affiliation[NIMS2]
{International Center for Materials Nanoarchitectonics, National Institute for Materials Science,  1-1 Namiki, Tsukuba 305-0044, Japan}
\author{Kausik Majumdar}
\email{kausikm@iisc.ac.in}
\affiliation[IISc]
{Department of Electrical Communication Engineering, Indian Institute of Science, Bangalore 560012, India}

\title{A High-Quality Entropy Source Using van der Waals Heterojunction for True Random Number Generation}


\begin{document}
\begin{abstract}
Generators of random sequences used in high-end applications such as cryptography rely on entropy sources for their indeterminism. Physical processes governed by the laws of quantum mechanics are excellent sources of entropy available in nature. However, extracting enough entropy from such systems for generating truly random sequences is challenging while maintaining the feasibility of the extraction procedure for real-world applications. Here, we present a compact and an all-electronic van der Waals (vdW) heterostructure-based device capable of detecting discrete charge fluctuations for extracting entropy from physical processes and use it for the generation of independent and identically distributed (IID) true random sequences. We extract a record high value ($>~0.98~bits/bit$) of min-entropy using the proposed scheme. We demonstrate an entropy generation rate tunable over multiple orders of magnitude and show the persistence of the underlying physical process for temperatures ranging from cryogenic to ambient conditions. We verify the random nature of the generated sequences using tests such as NIST SP 800-90B standard and other statistical measures and verify the suitability our random sequence for cryptographic applications using the NIST SP 800-22 standard. The generated random sequences are then used in implementing various randomized algorithms without any preconditioning steps.
\end{abstract}
\newpage

Random numbers find widespread use in everyday applications such as digital currencies, data encryption, etc., where the strength of the cryptographic protocol employed depends on the quality of the random number generator used\cite{Aljohani19}. They are also crucial in applications such as Monte Carlo methods for many-body simulations, statistical modeling, etc., and in a plethora of other probabilistic algorithms. The quality of the random number generator used influences the performance of these algorithms\cite{Timothy11}. Computers, being deterministic by design, are capable only of generating pseudo-random numbers in the absence of access to any physical entropy source. For applications like cryptography, this, along with the dramatic increase in the computational power seen over the past few decades, renders the system vulnerable to cryptanalysis. This has triggered a lot of research in designing true random number generators (TRNGs), both classical and quantum, which extract randomness from physical processes in nature. Quantum TRNGs rely on the probabilistic nature of the quantum events for randomness, while classical TRNGs such as TRNGs based on classical chaos rely on the indeterminism caused by finite measurement accuracy and the high sensitivity to initial conditions as the source of randomness. Utilizing these processes, TRNGs based on chaotic lasers\cite{Yoshiya20}, multi-modal ring oscillators\cite{Serrano21}, random Raman fiber lasers\cite{Monet21}, memristors\cite{Xuehua21,Kim21,Jiang17}, amplified spontaneous emission\cite{Liu13}, photon arrival time measurements\cite{He16}, superparamagnetic tunnel junctions\cite{Vodenicarevic17}, carbon nanotube transistors\cite{Sungho21}, etc. have been demonstrated recently. Together with the quality of randomness produced, the compactness and scalability of these technologies also play a crucial role in determining the real-life applicability of these physical processes as TRNGs. Optical processes, while being able to generate very high bit rates, are limited by the scalability and compactness they can offer. Photonic integrated circuits\cite{Ugajin17} could help circumvent this. TRNGs relying on faults or degradation of the system as the source of randomness have questionable reliability and stability over elongated periods of usage. This calls for the need for compact TRNGs based on physical processes which are stable over long periods of operations and are insensitive to their environment.

Owing to the criticality of applications that employ TRNGs and the diversity of physical processes used, a rigorous methodology to characterize the quality of randomness extracted by a TRNG is of utmost significance. A battery of tests as recommended in the National Institute of Standards and Technology (NIST) SP 800-22 statistical test suite\cite{Bassham10} are typically employed in literature to assess the randomness of the generated output bits\cite{Monet21,Xuehua21,Liu13,Sungho21,Ugajin17,Vodenicarevic17,He16}. We wish to emphasize that the NIST SP 800-22 battery of tests does not scrutinize the true randomness but instead analyzes the output of a true/pseudo-random source for features desirable for applications in cryptography. On the other hand, entropy has been in use in various fields of study as a measure to quantify the degree of uncertainty present in a system and was introduced to information theory by Shannon in his seminal paper\cite{Shannon48}. Entropy has also gained popularity as a measure to analyze different TRNGs for the quality of randomness produced\cite{Nie15,Corron16,Ugajin17,Tang15}. \txr{A pseudo-random source that is inherently deterministic contains zero entropy, whereas a true-random source contains high entropy. For true-random sources, the quality of randomness is characterized by the amount of entropy present. As an example, a biased coin toss, even though random, generates a lesser amount of entropy compared to that produced by an unbiased coin toss. Thus, maximizing the entropy is critical in ensuring that the source is free of any biases and hidden determinism.} To be able to maximize the entropy extracted and to utilize it for cryptographic applications, the random variable needs to be IID. Non-IID outputs require further conditioning steps, which could reduce the entropy, to be usable in many cryptographic protocols. To quantify the entropy extracted by a system and to assess the IID or non-IID nature of the generated output, a new standard NIST SP 800-90B\cite{Turan18} is currently in use. It focuses on analyzing the min-entropy of the system, which is a conservative estimate for the entropy extractable from the system and acts as an excellent measure to compare different sources of entropy.

In this work, we design and experimentally realize an all-electronic and highly scalable TRNG based on a black phosphorus (BP)/1L-WS\2/1L-MoS\2/1L-WS\2/graphene (Gr) van der Waals (vdW) heterostructure and generate an IID random output stream. The device relies on processes that are highly reliable, ensuring repeatability while maintaining a small footprint. The generated output passes the NIST SP 800-90B as well as NIST SP 800-22 test suites. We report an extracted min-entropy greater than $0.98~bits/bit$ and assess the operation of the heterostructure from cryogenic to ambient conditions. The high quality of the random sequence generated by the TRNG also enables us to demonstrate its use in multiple real-world applications without the need for any preconditioning steps.

\section{Results and Discussion}
\noindent\textbf{Experimental implementation and basic characterization.}

\begin{figure}[!h]
  \centering
  \includegraphics[width=16.00cm]{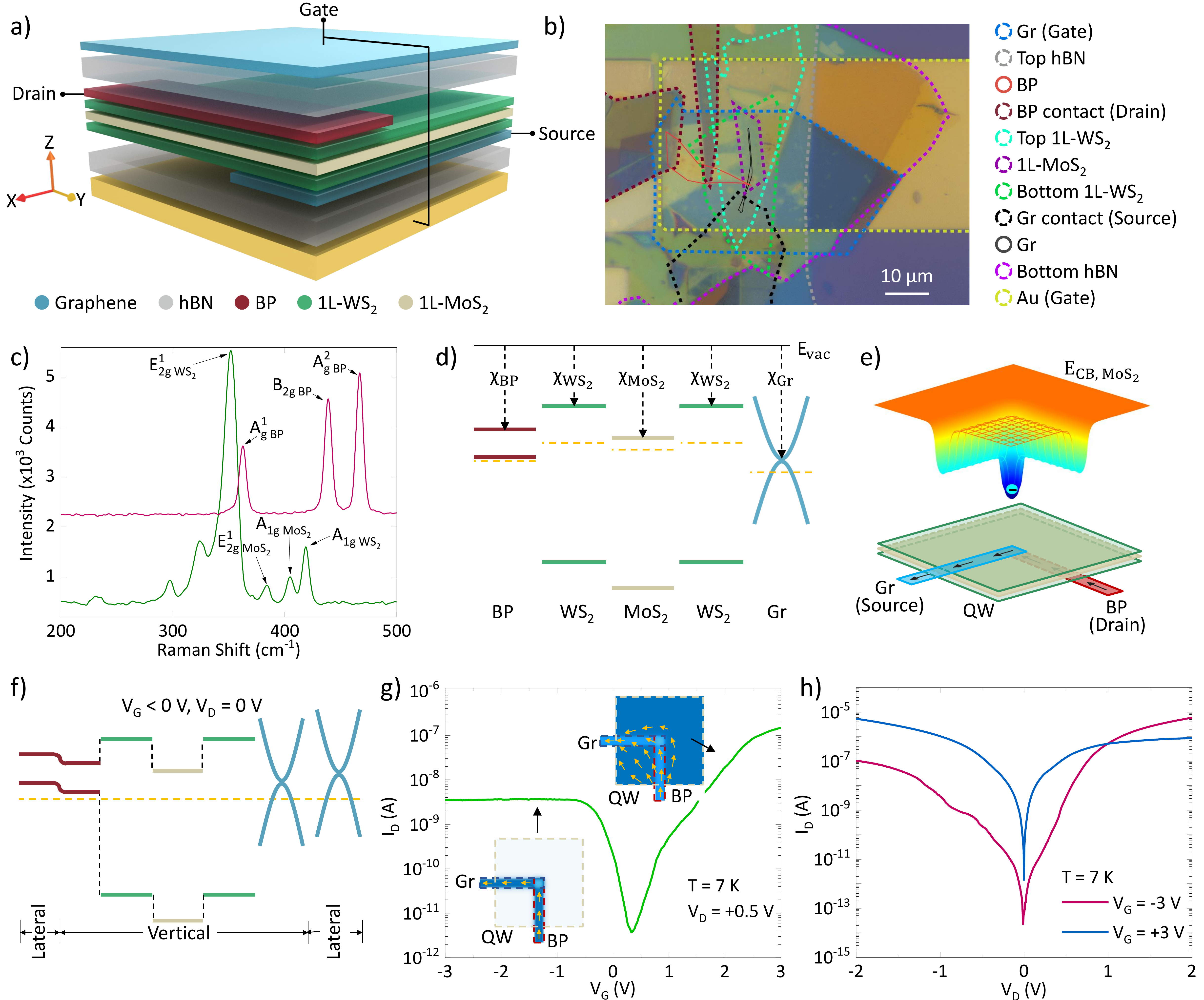}
  \caption{\label{fig:fig1}\textbf{Device structure and basic characterization.} \textbf{a)} Schematic representation of the device stack. Overlap between the BP and graphene layers define the quantum well in the x-y plane. \textbf{b)} A sample optical image of the device delineating various layers. \textbf{c)} Raman spectra of 1L-MoS\2, 1L-WS\2 and BP collected under a 532 nm excitation. \textbf{d)} Electron affinities, bandgaps and chemical potentials of the different layers before forming the heterostructure where $\chi_{BP}=4.2~eV,~\chi_{WS_2}=3.9~eV,~\chi_{MoS_2}=4.3~eV$ and $\chi_{Gr}=4.56~eV$. The orange dashed line indicate the Fermi level. \textbf{e)} A schematic profile of the MoS\2 CB under a negative $V_G$ depicting a strong confining potential for electrons at the junction. QW represents the WS\2/MoS\2/WS\2 stack. The device structure is outlined at the bottom with the arrows depicting the current flow under p-type operation. \textbf{f)} Band alignment of the heterostructure along the current flow path under a large negative gate bias. The orange dashed line indicates the Fermi level. \textbf{g)} $I_D-V_G$ characterization of the device A under a $V_D$ of $0.5~V$ at $7~K$. Insets show the different paths taken by the current under p-type (left) and n-type (right) operations. \textbf{h)} $I_D-V_D$ characterization of the device under large positive (blue trace) and negative (pink trace) gate biases.}
\end{figure}

The architecture of the TRNG is schematically depicted in Figure \ref{fig:fig1}a. The heterostructure consists of a 1L-WS\2/1L-MoS\2/1L-WS\2 stack sandwiched between a $\sim15~nm$ thick BP flake at the top and a few-layer Gr flake at the bottom. The stack is gated from the top using a Gr contact and from the bottom by an Au contact through hBN dielectric layers. Details of the device fabrication are given in \textbf{Methods}. A representative optical image of the final device and the Raman characterization of the flakes and are given in Figures \ref{fig:fig1}b and \ref{fig:fig1}c, respectively. Four such devices (A, B, C, D) are fabricated and characterized. The BP contact is referred to as the drain and the Gr contact as the source throughout the rest of the text. The top and bottom gate contacts are electrically shorted, forming a single gate terminal.

The band alignment of the constituent layers before forming the heterostructure is given in Figure \ref{fig:fig1}b with $\chi_{BP}=4.2~eV,~\chi_{WS_2}=3.9~eV,~\chi_{MoS_2}=4.3~eV$, and $\chi_{Gr}=4.56~eV$\cite{Abraham20,Frisenda18}. The alignment of the band extrema after forming the heterostructure is shown in Figure S1. The position of the 1L-MoS\2 conduction band (CB) relative to the CB of BP is verified by inspecting the sign of open-circuit voltage generated for a BP/1L-MoS\2 stack under a $1550~nm$ laser illumination, as shown in Figure S2. The large conduction band offset between 1L-WS\2 and 1L-MoS\2 helps us in creating an electron quantum well along the z-direction by sandwiching a 1L-MoS\2 layer between two 1L-WS\2 flakes. On the application of a large negative gate voltage ($V_G$), the MoS\2 and WS\2 layers are depleted of carriers and offer negligible screening to the gate electric field, whereas the BP and Gr layers become hole rich. This results in the MoS\2 region between the BP and Gr layers experiencing a lower gate electric field compared to the regions outside of the junction. This creates an electrostatically generated confinement for electrons in MoS\2 in the x-y plane, as shown in Figure \ref{fig:fig1}c. This effect, along with the quantum well formed along the z-direction by band engineering, localizes electrons in the MoS\2 layer of the junction. As the overlap between BP and Gr is the factor which determines the dimensions of the confining potential in the x-y plane, controlling the overlap area during fabrication is critical. We fabricate the devices with an overlap area ranging from $\sim 0.1~\mu m^2$ to $\sim 1~\mu m^2$. The energy band alignment along the direction of the current flow under a large negative $V_G$ is depicted in Figure \ref{fig:fig1}d. The lateral regions of both BP and Gr are of stronger p-type doped compared with the junction as a result of being gated by both the top and bottom gates. This ensures that the electrical characteristics of the device are dominated by the junction by minimizing the effects of undesired series resistance. Thus, the careful design of the heterostructure enables us to utilize the varying extent of screening at different regions of the device to generate the desired confining potential, all the while employing a simple gating arrangement.


We measure the modulation of the drain current ($I_D$) with $V_G$ at a fixed drain voltage ($V_D$) of $+0.5~V$. The obtained curve for device A at a temperature of $7~K$ is given in Figure \ref{fig:fig1}e. For increasing positive values of $V_G$, all the constituent layers become n-doped. As the WS\2 and MoS\2 layers are also electron-rich, the current transport is not limited to through the junction but takes multiple pathways from drain to source (Figure \ref{fig:fig1}e inset). Hence, $I_D$ exhibits an increasing trend with increasing $V_G$. For the case of negative $V_G$, the WS\2 and MoS\2 layers are depleted of carriers, and the current transport is forced only through the tunnel junction (Figure \ref{fig:fig1}e inset). Holes injected from the drain contact into the BP valence band (VB) drifts through the BP layer towards the junction, where it tunnels into Gr through the barrier created by the VB offset of the 1L-WS\2/1L-MoS\2/1L-WS\2 stack. The effect of the gate field screening at the junction is clearly visible in the $I_D-V_G$ curve at larger negative values of $V_G$ as a flattening of the curve. The observed gate leakage current from the device is given in Figure S3, where ultra-low leakage current rules out any saturation effects occurring as a result of gate leakage. The measured $I_D-V_D$ characteristics of the same device under large positive and negative $V_G$ are shown in Figure \ref{fig:fig1}f. For small values of $V_D$, the n-type current is few orders of magnitude higher than the p-type current as a result of limiting the current flow through the junction for p-type operation (Figure \ref{fig:fig1}e inset). For larger values of $V_D$, the comparison is not credible as the saturation effects due to the changing $V_G$-$V_D$ start contributing to the observed characteristics. Under p-type operation, as the entirety of $I_D$ is controlled by the tunneling resistance offered by the 1L-WS\2/1L-MoS\2/1L-WS\2 hole barrier, $I_D$ behaves as a suitable probe for measuring the changes in the junction, which will form the foundation for the operation of the TRNG.\\

\noindent\textbf{Dynamics of the TRNG device.}

\begin{figure}[ht]
  \centering
  \includegraphics[width=16.25cm]{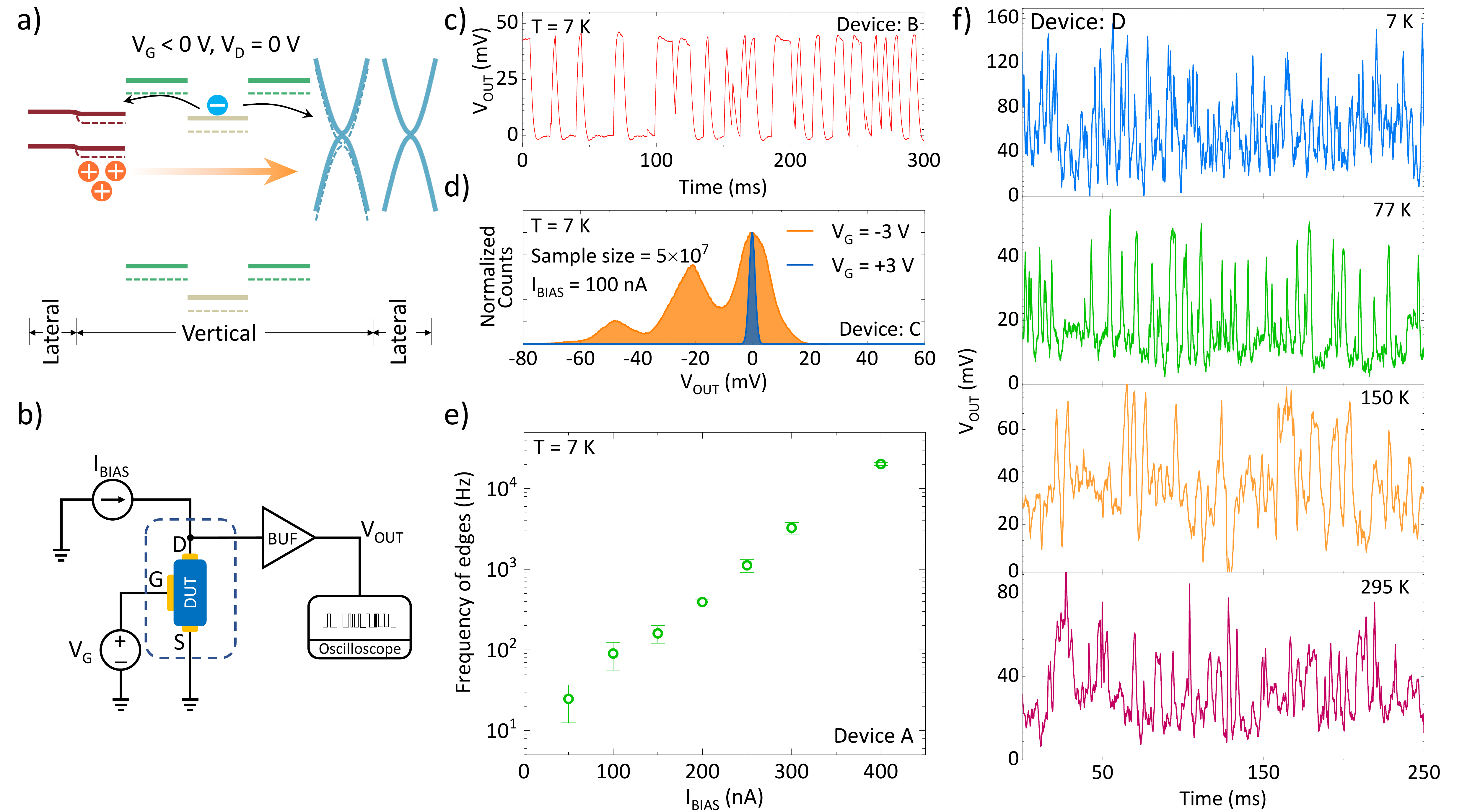}
  \caption{\label{fig:fig2}\textbf{Principle of operation, temperature dependence, and frequency modulation.} \textbf{a)} Modulation of the energy band alignment of the heterostructure with the trapping of an electron in the conduction band of the quantum well. Black arrows indicate the possible de-trapping pathways and the orange arrow indicates the hole current under a positive $V_D$. \textbf{b)} Illustration of the measurement setup used for data acquisition. BUF is a unity gain voltage follower. Elements depicted inside the blue dashed line are inside a cryostation. \textbf{c)} A sample voltage vs. time serial data from device B, as obtained from the oscilloscope. \textbf{d)} Normalized histograms of output voltages from device C, at $V_G=+3~V$ (blue) and $V_G=-3~V$ (orange) biases showing stark differences in the output behavior obtained from $50$ million $V_{OUT}$ data points per $V_G$. The $V_{OUT}$ values are shifted to match the highest resistance state with $0~V$. \textbf{e)} Modulation of the frequency of edges (rising and falling) in $V_{OUT}$ as a function of $I_{BIAS}$ from device A, showing exponential growth. Error bars correspond to $\pm$ 1 standard deviation. \textbf{f)} Raw voltage vs time data from device D for various operating temperatures ranging from 7 K (top panel) to 295 K (bottom panel) showing similar voltage spikes at an $I_{BIAS}$ of $20~nA$.}
\end{figure}

As the overlap area decreases, the MoS\2 region in the junction acts as an electrostatically created island for electrons, and discrete charge effects become prominent (Figure \ref{fig:fig1}c). Although completely depleted of carriers by the negative $V_G$, stray electrons in the vicinity of the junction can funnel into the island and get trapped in the CB of MoS\2. The trapped electron affects the electrostatics of the device mainly in two ways (see Figure \ref{fig:fig2}a). First, the added negative charge in the junction acts as a local gate and raises the energy bands of the nearby regions by Coulomb interaction. This lowers the hole barrier along the in-plane direction created as a result of screening of gate fields and thereby increases $I_D$. Second, as the electron trap and the hole tunnel barrier are created by the same stack, the charging of the electron trap increases the potential energy of the trap\cite{Averin86}. This lowers the hole tunnel barrier and causes $I_D$ to increase. The trapped electron can escape the island either by tunneling into the Gr layer or by overcoming the small barrier provided by BP CB, as shown by the arrows in Figure \ref{fig:fig2}a. Detrapping into the Gr layer is expected to be faster than escaping by thermionic field emission into BP CB. However, for the rest of the text, we operate the device under a positive $V_D$. The field created by a positive $V_D$ suppresses the fast decay into Gr and decay via charge transfer into BP dominates. This is done to slow down the de-trapping events so that such events can be accurately captured in the experimental setup with limited bandwidth.

To analyze the transient nature of the capture and release of electrons by the island, we employ the setup shown in Figure \ref{fig:fig2}b. Details of the measurement procedure are given in \textbf{Methods}. The measured time trace from device B at $7~K$ is given in Figure \ref{fig:fig2}c. The rising (falling) edges in the measured $V_{OUT}$ correspond to an increase (decrease) in the device resistance as a result of electron release (capture) from the island. The trapping and de-trapping dynamics of the electron by the island is clearly visible for all the fabricated devices, although the devices with larger area tend to exhibit a smaller spike amplitude as a result of reduced trapped carrier density. The mechanism of operation of our device shares some similarities with the random telegraph noise observed in MOSFETs\cite{Brown18}, but differs by employing CB states to capture and store electrons. Although it is possible for some of the electrons stored in the CB to fall into the deep traps in MoS\2\cite{Furchi14}, the device operation is not adversely affected. This, along with the different capturing and escaping dynamics, helps the device maintain versatility in operating conditions while ensuring reliability. We observe similar dynamics in all the fabricated devices, suggesting a high throughput for the proposed device architecture. The reliable operation of the device is a result of choosing highly robust physical processes which do not rely on any process-dependent variability or the presence of defects as a source of randomness. \txr{By employing CB levels as the basis of the electrostatically created island, we avoid dependence of defect based charge trapping which enables faster dynamics as well as repeatable behaviour. Further, the interaction between the trapped electron and the probe hole current $I_D$ being electrostatic in nature, strengthens the reliability of the device operation.}

For a comprehensive analysis of the device, it is critical to inspect the contribution from other sources of noise such as shot noise or thermal noise, and to rule out any setup-induced errors. In this regard, we measure the output of the device at p-type and n-type operation with a fixed $I_{BIAS}$ of $100~nA$ at $7~K$. Obtained histograms of $V_{OUT}$ under $V_G=-3~V$ and $V_G=+3~V$ are given in Figure \ref{fig:fig2}d. The measured $V_{OUT}$ values are offset by a DC value to match the highest resistance state with $0~V$. As expected, the distribution of $V_{OUT}$ exhibits a single narrow peak for $V_G=+3~V$. On the contrary, for $V_G=-3~V$, the histogram features multiple well-separated peaks suggesting an origin different from the usual sources of noise. The different peaks correspond to the varying number of captured electrons in the island, which in turn results in multiple discrete values for the device resistance.\\

\noindent\textbf{Tunability of entropy generation rate.}

We assess the dependence of the rate of events on the magnitude of $I_{BIAS}$. With increasing values of $I_{BIAS}$, the capturing and releasing rates of electrons by the island increase. For a positive $I_{BIAS}$, the quantum well floats to a positive potential determined by the coupled Poisson and continuity equations. This results in a stronger field for electrons in the vicinity of the junction to drift into the island, thereby increasing the capturing rate. The increased $I_{BIAS}$ could also have local heating effects which can lead to an increase in the generation rate of stray electrons. As the main pathway for electrons to escape the island is to tunnel into the BP CB, the increasing field created by the rising $I_{BIAS}$ results in a higher de-trapping rate as well. The observed frequency of changes in $V_{OUT}$ caused by trapping and de-trapping of electrons, as a function of $I_{BIAS}$ for our TRNG is given in Figure \ref{fig:fig2}e showing exponential growth with increasing values of $I_{BIAS}$. The exponential dependence suggests that recombination of captured electrons with holes, although possible, might have a lower contribution to the escape rate, and the curve could be dominated by changes in the generation and de-trapping rates by $I_{BIAS}$. We do not observe any degradation in the quality of randomness with an exponential increase in rate as will be demonstrated later in the text after introducing the necessary tools for analysis. The high tunability in the frequency of events enables the TRNG to operate at a broad range of entropy generation rates as required by the end application. \txr{For non-critical applications such as gaming or when only periodic reseeding is sufficient, the system might be able to operate at a low entropy feed from the TRNG and hence, operate at a lower power. For security-critical applications such as cryptographic key generation that require a high entropy feed from the TRNG, the device could be switched to a faster operation by increasing the $I_{BIAS}$ and thereby resulting in a high entropy generation rate at the expense of higher operating power. In such scenarios, having the ability to tune the entropy generation rate could be beneficial in dynamically controlling and minimizing the power consumption of the TRNG while retaining reliable operation.} The $I_{BIAS}$ could also be used as a tunable parameter to stabilize the entropy generation rate of the TRNG against changes in the operating temperature of the device.\\

\noindent\textbf{Temperature dependence of the TRNG operation.}

The stability of the TRNG against changes in its environment is conditional on the stability of the physical process generating the entropy. A wide range of operating temperatures is a desirable characteristic for a TRNG. Due to the increasing demand for cryogenic circuits, TRNGs that support low-temperature operations as well are beneficial. Now we examine the temperature dependence of the underlying physical process. The measured time trace of $V_{OUT}$ for temperatures ranging from $T=7$ to $295~K$ is given in Figure \ref{fig:fig2}f. The voltage pulses generated by the capture and release of electrons by the island are evident in the measured data indicating the applicability of the physical process as a good entropy source for a wide range of temperatures. This is made possible by the large bandgaps offered by the WS\2 and MoS\2 monolayers enabling us to deplete them of carriers even at higher temperatures. The nature and randomness of the stream generated by the device at $295~K$ remains stable without any degradation from the response at $7~K$ as will be shown after detailing the steps to extract a random stream from our TRNG. The coupling mechanism between the trapped electron and the probing hole current being electrostatic in nature also remains largely unaffected by the change in the temperature. The parameters which are expected to show temperature dependence are the de-trapping rate determined by the thermionic field emission and the generation rate of stray electrons. Both the rates are expected to increase with an increase in temperature, thereby increasing the frequency of the events. Contrary to the expectation, the data does not exhibit any enhancement in the frequency of events at higher temperatures which we believe is a result of the stray capacitances of the experimental setup leading to averaging out of the fast pulses. This can be avoided by implementing the voltage buffer used in the setup in the same chip as the TRNG device.\\

\noindent\textbf{Choice of random variable.}

\begin{figure}[ht]
  \centering
  \includegraphics[width=16.5cm]{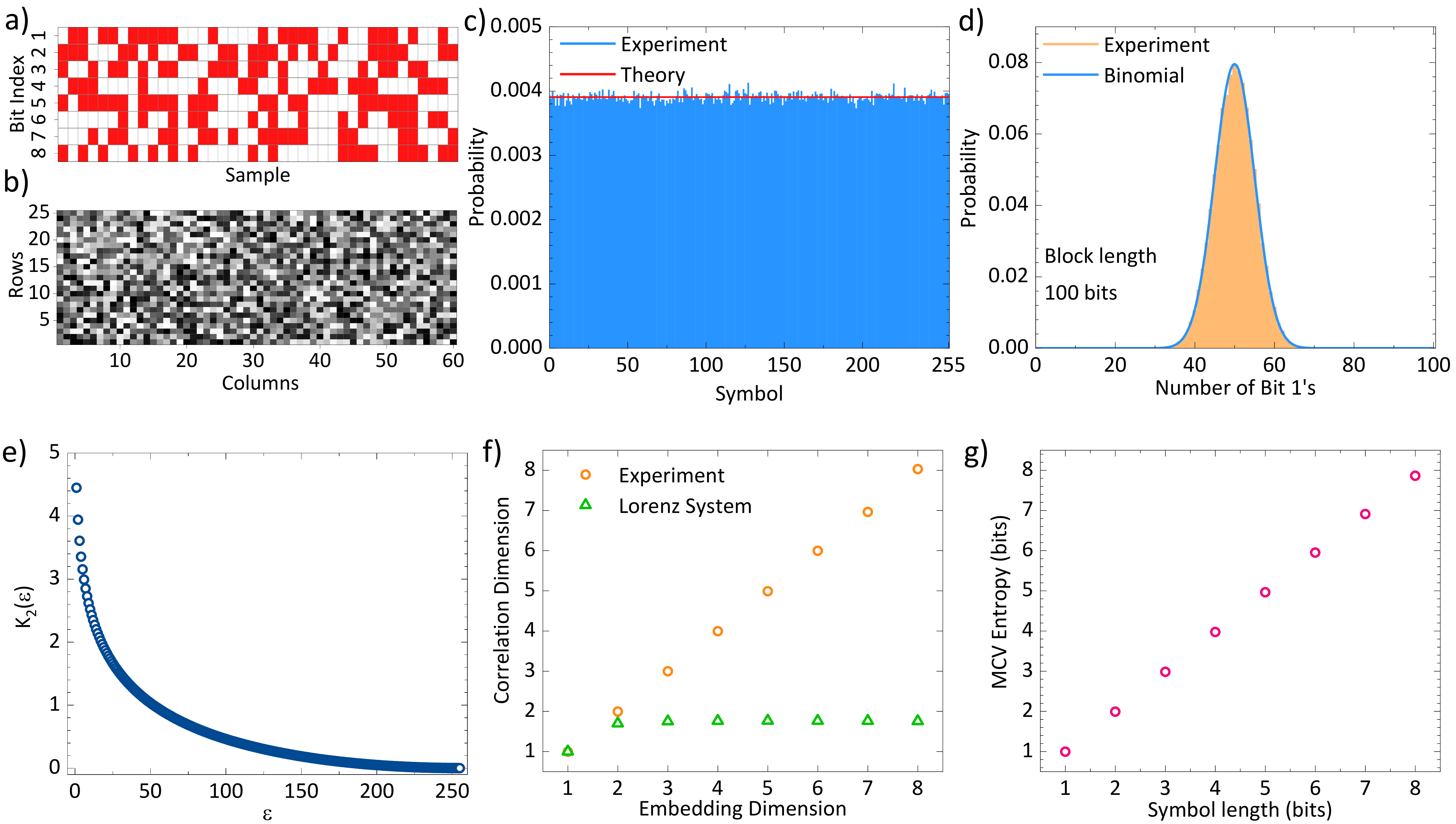}
  \caption{\label{fig:fig3}\textbf{Examining randomness of the TRNG output from device D.} \textbf{a)} A slice of the output stream obtained after digitizing the time bins with 8 bit resolution. White (red) squares represent bit 1 (bit 0).\textbf{b)} Grayscale map obtained by reshaping a stream of 1500 samples into a $25\times60$ matrix. 8 bit values ranging from 0 to 255 are mapped to the intensity of the pixel, with 255 corresponding to white. \textbf{c)} Probability distribution for all the symbols in the alphabet obtained experimentally (blue bars) and that of a theoretical uniform distribution (red trace). \textbf{d)} Probability distribution of the number of bit 1's in a binary block of length 100 bits. Experimental (orange bars) closely follow a theoretical unbiased binomial distribution (blue trace). \textbf{e)} Divergence of the K\2 entropy as a function of decreasing distance threshold $\varepsilon$. \textbf{f)} Correlation dimension $\nu$ estimate as a function of the embedding dimension $d$ showing a linear relation with a unity slope for the experimental data (circle markers) and a saturating behavior for a chaotic Lorenz system (triangle markers). \textbf{g)} Estimated MCV entropy of the data for various digitizer bit lengths showing a linear trend.}
\end{figure}

Due to the quantum mechanical nature of the discrete single-electron processes involved (like generation of stray electrons followed by their funnelling into the island and out-tunneling to contacts), the time points of the capture and release events are random in nature. Thus we choose the indices of the time bins at which the edges (rising and falling) of $V_{OUT}$ occur as the random variable. \txr{Relying on the time instants of these events rather than the amplitude fluctuations in $V_{OUT}$ also provides tolerance to external noise and the device dimensions.} The time axis is divided into repeating blocks, each with $k=2^n$ bins where $n$ is the width of the symbol $x_i$ outputted by the random variable. More details on the generation of the random sequence from the device output is given in \textbf{Methods}. The set of all possible symbols $\{x_1,x_2,...,x_k\}$ is called the alphabet on which the TRNG operates. To ensure the independence of the output of the TRNG, the sampling rate of the time trace should be high enough so that the probability of more than one event occurring in a block is negligible. Hence the sampling rate and the maximum possible alphabet size for the TRNG are interrelated. Here we employ an alphabet of $k=256$ symbols corresponding to an $n=8~bit$ wide symbol. A slice of the random sequence generated by the TRNG (device D biased at an $I_{BIAS}=20~nA$ and sampled at $12.5~MS/s$) is shown in Figure \ref{fig:fig3}a.\\

\noindent\textbf{Randomness analysis using statistical measures.}

We employ different methodologies of varying complexity to assess the quality of the random sequence generated by the device. Figure \ref{fig:fig3}b shows a grayscale map generated from a slice of 1500 samples of $8~bit$ width extracted from the output stream of the TRNG operated at $7~K$. The absence of any clear pattern provides a preliminary prediction towards the possible random nature of the output sequence. The average amount of information generated by the TRNG at each output, also known as the Shannon entropy\cite{Shannon48}, is directly correlated with the degree of randomness of the system generating the output and is given by $H_S=-\sum_{i=1}^kP(x_i)~log_2P(x_i)$, where $P(x_i)$ is the probability for observing $x_i$. For sources that produce deterministic outputs, the Shannon entropy is $0~bits/bit$, whereas it is $1~bit/bit$ for completely uncertain events. $H_S$ is maximized when $x_i$ is identically distributed. Many of the TRNGs fail to produce an IID output, thereby limiting the maximum extractable entropy. In Figure \ref{fig:fig3}c, we measure the probability distribution of the various symbols from the output data. The experimentally obtained data (blue bars) match well with the theoretical distribution (red line) for a uniform random variable. This helps us in achieving an $H_S$ of $7.9998~bits/symbol$ or $0.9999~bits/bit$ for our TRNG.

The identical distribution of the symbols is also evident from the distribution of $bit~0s$ and $bit~1s$ in the generated output bitstream. In Figure \ref{fig:fig3}d, we plot the distribution of $bit~1s$ in $100~bits$ wide blocks in the output bitstream. The measured distribution (orange bars) matches well with the expected binomial distribution (blue line) for an unbiased system. This indicates that each bit of the output symbol has equal probabilities in taking a value of 0 or 1. The binomial nature of the distribution of $bit~1s$ in a finite block is preserved even when the entropy generation rate is increased by a few orders in magnitude by a modulation in $I_{BIAS}$, as is shown in Figure S4. This enables the $H_S$ to be virtually unaffected by the increase in the rate of events (Figure S5), proving the high tunability offered. We demonstrate an entropy generation rate of $\sim 162~kb/s$ (operating at an energy of $\sim 2.3~pJ/bit~of~entropy$) and should emphasize that the $H_S$ or the rate of events as shown in Figure \ref{fig:fig2}e does not show any saturation with $I_{BIAS}$, hinting at a possible higher intrinsic entropy generation rate of our architecture.

The analysis is repeated for the TRNG operation at $295~K$ to assess the nature of the device operation outside cryogenic conditions. The generated bitsream and a grayscale map generated from room temperature operation are given in Figures S6a and S6b showing no obvious patterns. All the various symbols of the alphabet also are generated with equal probability as is evident from Figure S6c enabling the TRNG to achieve an $H_S$ of $0.9993~bits/bit$ at $295~K$. The distribution of $bit~1s$ also exhibits a binomial distribution similar to operation at $7~K$ (Figure S6d) proving the stablility of the TRNG operation in temperatures from $7~K$ to $295~K$.

Analyzing the distribution of symbols, although important, is not sufficient to ascertain the true randomness of the sequence. Therefore, we make use of methods typically employed in the study of chaotic dynamics in complex nonlinear systems to probe for the presence of any hidden determinism in the device operation. In this regard, Kolmogorov entropy\cite{Kolmogorov58} ($K$), which quantifies the mean rate of information generated by a system, is a good metric. Here, the term $information$, although conceptually related to $Shannon~information$, is not mathematically equivalent\cite{Falniowski14}. $K$ is related to the predictability time $\tau$ as $K\sim \tau^{-1}$\cite{Cohen85}. $\tau$ estimates the average time into the future for which we can predict the evolution of the system based on the knowledge about its history. Therefore, a non-chaotic deterministic system has $K=0$, a finite value for $K$ for a chaotic system, and $K=+\infty$ for a random process\cite{Grassberger83,Cohen85}. Estimation of $K$, although easy for analytical models, is difficult for experimental data. To circumvent this, another quantity, $K_2$\cite{Grassberger83}, which is a lower bound for $K$, is typically employed. Details on estimating $K$ and $K2$ entropies is given in \textbf{Methods}. The calculated values of $K_2(\varepsilon)$ for the experimental data are given in Figure \ref{fig:fig3}e and show a clear divergence for decreasing distance threshold $\varepsilon$ ascertaining the stochastic nature of the TRNG output. The absence of correlations between the outputs of the TRNG is also evident from the absence of any discernible pattern in the difference and lag plots generated from the output stream given respectively in Figures S8 and S9.

To supplement the analysis based on $K_2$, we also probe the system for the presence of any underlying low dimensional dynamics by estimating the growth of correlation dimension $\nu$ of the system with the embedding dimension $d$ used. Details on estimating $\nu$ is given in \textbf{Methods}. The estimated $\nu~vs~d$ plot for the experimental data is shown in Figure \ref{fig:fig3}f along with that for a chaotic Lorenz system\cite{Lorenz63} showing a non-saturating trend for the TRNG output as expected for a stochastic system.\\

\noindent\textbf{Evaluation against standardized protocols.}

Once the device is shown to be a source of true randomness, we run the NIST recommended standard, SP 800-90B test suite\cite{Turan18} on the output of the TRNG to evaluate the quality of the randomness produced by the system and to assess its suitability for high-end applications such as cryptography. The requirements on the data collection and the various steps in assessing the entropy source against the SP 800-90B standard are detailed in the \textbf{Methods} section. \txr{A sequential and restart datasets of 8 million bits each are collected and analyzed. The results from these tests for the generated dataset are shown in Tables \ref{tab:tab1} and \ref{tab:tab2}.} The experimentally generated random sequence passes all the tests certifying its IID nature as presumed on the basis of the previous analyses. The extracted min-entropy of the IID output of our TRNG as a function of varying time-bin bit width $n$ is given in Figure \ref{fig:fig3}g. We report a min-entropy greater than $0.98~bits/bit$ for all $n$ ranging from 1 to 8. To the best of our knowledge, this is the highest reported min-entropy for any TRNG. \txr{Such high min-entropy extracted by SP 800-90B further strengthens the high quality and truly random claims for the bit-stream generated by the TRNG.} An increase in $n$ and thereby the total entropy generated can be achieved either by increasing the timing resolution or by adding multiple devices in parallel. \txr{These processes could still be used to increase the total entropy even when the $I_{BIAS}$ has reached the maximum permissible values set by the material properties.} Raising the timing resolution is not a scalable option, whereas the all-electronic nature of the device architecture enables us to pack a large number of devices without much overhead. Such scalability in the architecture is essential for the advancement of the technology towards practical applications.

\begin{table}
\centering
\begin{tabular}{|l|c|c|c|}
\hline
\multirow{2}{*}{\textbf{Test}}&\multirow{2}{*}{\textbf{Sequential}}&\multicolumn{2}{|c|}{\textbf{Restart}}\\\cline{3-4}
& & \textbf{Rows}&\textbf{Columns}\\
\hline
Excursion Test Statistic & Pass & Pass & Pass \\\hline
Number of Directional Runs & Pass & Pass & Pass \\\hline
Length of Directional Runs & Pass & Pass & Pass \\\hline
Number of Increases and Decreases & Pass & Pass & Pass \\\hline
Number of Runs Based on the Median & Pass & Pass & Pass \\\hline
Length of Runs Based on Median & Pass & Pass & Pass \\\hline
Average Collision Test Statistic & Pass & Pass & Pass \\\hline
Maximum Collision Test Statistic & Pass & Pass & Pass \\\hline
Periodicity Test Statistic (Lag=1) & Pass & Pass & Pass \\\hline
Periodicity Test Statistic (Lag=2) & Pass & Pass & Pass \\\hline
Periodicity Test Statistic (Lag=8) & Pass & Pass & Pass \\\hline
Periodicity Test Statistic (Lag=16) & Pass & Pass & Pass \\\hline
Periodicity Test Statistic (Lag=32) & Pass & Pass & Pass \\\hline
Covariance Test Statistic (Lag=1) & Pass & Pass & Pass \\\hline
Covariance Test Statistic (Lag=2) & Pass & Pass & Pass \\\hline
Covariance Test Statistic (Lag=8) & Pass & Pass & Pass \\\hline
Covariance Test Statistic (Lag=16) & Pass & Pass & Pass \\\hline
Covariance Test Statistic (Lag=32) & Pass & Pass & Pass \\\hline
Compression Test Statistic & Pass & Pass & Pass \\\hline
Chi-Square Independence & Pass & Pass & Pass \\\hline
Chi-Square Goodness of Fit & Pass & Pass & Pass \\\hline
Length of the Longest Repeated Substring & Pass & Pass & Pass \\\hline
\end{tabular}
\caption{\label{tab:tab1}Tests for IID assumption.}
\end{table}

\begin{table}
\centering
\begin{tabular}{|l|c|c|}
\hline
\multirow{2}{*}{\textbf{Dataset}}&\textbf{Sanity}&\textbf{Entropy}\\
&\textbf{check}&\textbf{estimate}\\
\hline
Sequential & NA & 7.8644 \\\hline
Restart Rows & Pass & 7.8640 \\\hline
Restart Columns & Pass & 7.8640 \\\hline
\multicolumn{3}{|c|}{$\textbf{min-entropy=7.8640~bits/8~bits}$}\\\hline
\end{tabular}
\caption{\label{tab:tab2}Restart tests.}

\end{table}

A reliable TRNG is expected to generate random streams of stable entropy throughout its operation. The system should be capable of detecting any departure from the expected behavior. To monitor the quality of the randomness on the fly, NSIT SP 800-90B requires a set of health tests to be run indefinitely on the TRNG output stream. The recommended tests are the repetition count test and the adaptive proportion test\cite{Turan18}. The repetition count test checks if any symbol is consecutively produced more often than predicted by the min-entropy. The adaptive proportion test checks if a randomly chosen symbol occurs too frequently in a finite block than expected. Our TRNG passes both the health tests during the course of its operation, results of which are given in Figures S11a and S11b, indicating a stable and reliable entropy generation.

Together with the high value for min-entropy, the IID nature of the output enables us to directly use the generated output stream for cryptographic applications without the need for any conditioning of the output. This is further verified by the results of NIST SP 800-22 statistical test suite results given in Table S1. To benchmark the performance of the TRNG demonstrated in this work, the proposed device is compared with other TRNGs in the literature\cite{Serrano21,Yingchun20,Seongmo19,Gehring21,Yoshiya20,Monet21,Gaviria17,Xuehua21,Liu13,Woo21,Brown18,Vodenicarevic17}, which employ different physical processes in various platforms for random number generation, the results of which are given in Table S2. \\

\noindent\textbf{TRNG usage in real-life applications.}

\begin{figure}[ht]
  \centering
  \includegraphics[width=16.5cm]{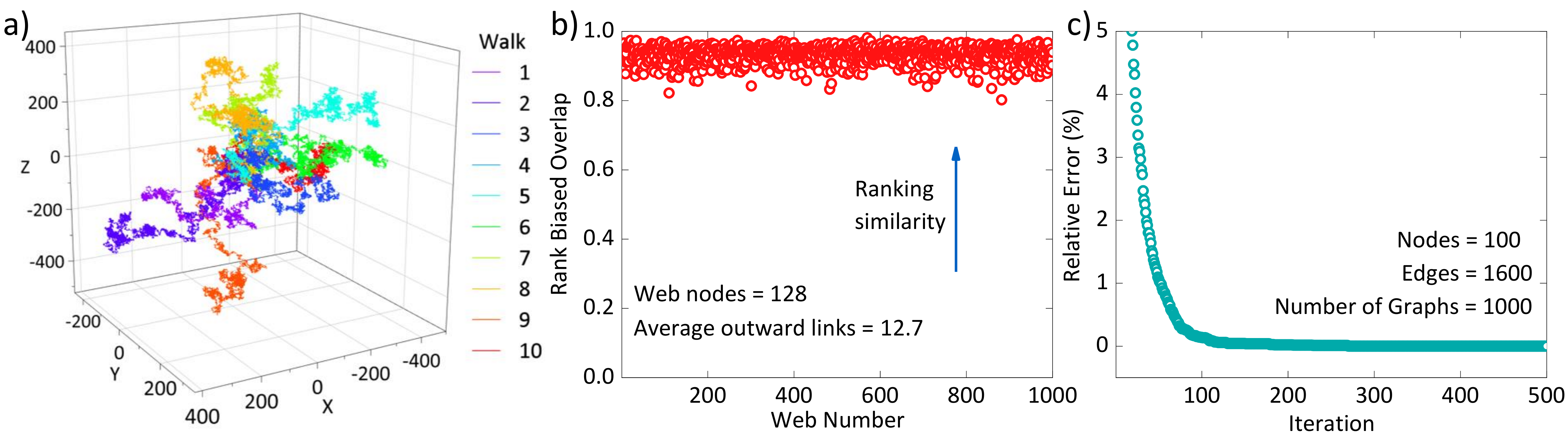}
  \caption{\label{fig:fig4}\textbf{Applications of the TRNG output.}  \textbf{a)} Random walks on an infinite cubic lattice generated from ten different sample streams. \textbf{b)} Similarity of PageRanks computed by power iteration and from random walks created with the experimental data on 1000 randomly generated webs. \textbf{c)} Convergence of the minimal cut estimate of 1000 randomly created undirected graphs to the true value for a randomized algorithm employing the experimentally generated random numbers.}
\end{figure}

In order to demonstrate the applicability of the generated random sequence in real-life scenarios, we employ our TRNG (device D) as the source of randomness in some of the well-known randomized algorithms. An important object typically encountered in randomized algorithms is random walk. Random walks are trajectories where each step of the trajectory is chosen at random from a set of possibilities. They are critical in modeling various phenomena in physics, genetics, ecology, brain research, etc\cite{fern92,Chipman09}. Few random walks created by our TRNG in an infinite cubic lattice are shown in Figure \ref{fig:fig4}a.

We employ random walks generated from the TRNG in assessing the PageRank\cite{Page99} of a highly connected web. PageRank is an algorithm developed by Google for ranking pages on the web depending upon the number of links directed to a node. On a randomly created web, we execute a randomized algorithm employing random walks created by the TRNG to estimate the ranks of each page on the web. The generated ranking is compared with the true ranks obtained using a power iteration method. The similarity between rankings is measured using rank biased overlap\cite{Webber10} (RBO), which estimates the similarity between two ranked lists with higher weights for top ranks. The obtained results for a set of 1000 webs are given in Figure \ref{fig:fig4}b showing the high degree of accuracy obtained by the randomized algorithm.

Another important use of randomized algorithms is in estimating the minimum cut of a graph which measures the minimum number of edges to be cut to make a graph disjoint. This is typically used in image segmentation tasks and in analyzing the vulnerability of various networks to the failure of links\cite{Boykov01}. For a graph with a large number of nodes, the time complexity of the problem grows fast, which renders deterministic algorithms difficult for real-world applications. Randomized solutions with much lesser time complexity exist for the minimum cut problem where the accuracy of the estimation grows rapidly with iteration\cite{Karger93}. In Figure \ref{fig:fig4}c, we show the accuracy of the results obtained from a randomized algorithm\cite{Karger93} based on random sequences generated by our TRNG as a function of the algorithm iteration. The error estimate is based on results from 1000 randomly generated graphs and approaches zero error rapidly, suggesting the high quality of random sequences generated by the TRNG.

\section{Conclusions}
In this work, we demonstrate a vdW heterostructure-based physical entropy source for generating true random numbers capable of extracting entropy close to the ideal limit. The compact and all-electronic nature can help in easy integration of the device with existing systems, providing them with access to a high-quality entropy source. Minimizing the stray capacitances in the design with on-chip elements (buffers, biasing circuitry, etc.) can help further enhance the large tunability in the entropy generation rate offered by the device. The IID character of the generated sequence and the lack of need for any preconditioning step makes the device a feasible solution to the requirements on the random number generators set by high-end applications such as cryptography.

\section{Methods}
\subsection{Fabrication}
The pre-patterned contacts are defined by optical lithography using a 360 nm UV source on an AZ5214E resist coated Si/SiO\2 wafer with 285 nm thick oxide formed by dry chlorinated thermal oxidation and forming gas annealing. A 20 nm thick Ti film followed by a 40 nm thick Au film is deposited by DC magnetron sputtering and lifted off using an acetone/isopropyl alcohol rinse to form the contacts. Flakes of the desired material are exfoliated from bulk crystal using Scotch tape and are transferred to a poly dimethyl-siloxane (PDMS) sheet. Flakes of suitable thickness are identified under an optical microscope by optical contrast and are dry transferred onto the substrate with precise control over the alignment. The final stack is then annealed at $200~\degree$C for 3 hours under $10^{-6}$ Torr pressure.
\subsection{Characterization}
The device is loaded into a closed cycle He probe station (Lakeshore CRX-6.5K) with viewports sealed to maintain a dark environment. The basic DC characterization and the required biasing during entropy extraction are done using a parameter analyzer (Keithley 4200A SCS). For the transient characterization, a constant current $I_{BIAS}$ is forced into the junction through the drain contact. The output signal containing the information about the capture and release dynamics of the electron in the form of voltage modulation is then fed through a unity gain non-inverting voltage buffer (Texas Instruments LMP7721MAEVALMF/NOPB) and is DC level shifted before sending to a digital storage oscilloscope (Tektronix MDO3000). All the measurements are done in a vacuum with a pressure $<1\times 10^{-4}$ Torr.
\subsection{Generation of the random variable}
The voltage signal $V_{OUT}$ is high-pass filtered to remove any DC bias as we are only interested in the time at which the events occur. The edges are then detected by thresholding the filtered signal. The output from this is used to identify the corresponding time bin (n-bit-wide). The index of the time bin (the time instant represented in an n-bit-wide alphabet) forms the output of the random variable and results in n random bits.

\subsection{$K$ and $K2$ entropy estimation}
For a system with $F$ degrees of freedom, $K$ is given by\cite{Grassberger83},
\begin{equation}
K=-\lim_{\tau\to0}\lim_{\epsilon\to0}\lim_{d\to\infty}\frac{1}{d\tau}\sum\limits_{i_1,...,i_d}p(i_1,...,i_d)ln[p(i_1,...,i_d)]
\end{equation}
where $\tau$ is the time resolution, $\epsilon^F$ is the volume of a hypercube in an $F$ dimensional phase space, and $i_j$ is the index of the hypercube in the partitioned phase space occupied by the system at time $j\tau$. A set of numbers $i_1,...,i_d$ defines a specific trajectory of $d$ steps for the system, and $p(i_1,...,i_d)$ is the joint probability. The summation is taken over all possible trajectories, and $ln$ is the natural logarithm.

$K_2$ is a special case of $q^{th}$ order R\'{enyi} entropy $K_q$ with $q=2$, where $K_q$ is given by\cite{Grassberger83}
\begin{equation}
K_q=-\lim_{\tau\to0}\lim_{\epsilon\to0}\lim_{d\to\infty}\frac{1}{d\tau}\frac{1}{q-1}ln\sum\limits_{i_1,...,i_d}p^q(i_1,...,i_d)
\end{equation}
and $0\leq K_2\leq K$. $K_2$ can be quickly estimated from recurrence plots for varying distance thresholds $\varepsilon$\cite{Faure98}. A recurrence plot is a graphical way of representing instants in the trajectory of the system, which occupies phase space regions within a distance threshold $\varepsilon$. A sample recurrence plot for the experimental data is given in Figure S7. As $K_2$ is upper bounded by $K$, by analyzing the behavior of $K_2(\varepsilon)$ for decreasing values of $\varepsilon$, we can make predictions about the nature of the dynamics of the system. For stochastic systems, $K_2(\varepsilon)$ diverges as $\varepsilon$ reduces\cite{Faure98}.

\subsection{Correlation dimension estimation}
To estimate the correlation dimension, the one-dimensional output stream is projected onto higher dimensions using time-delay embedding\cite{Weigend94} where $d$ values from the output separated by a suitable lag form a $d-$tuple in the $d$ dimensional space. $d$ forms the embedding dimension of the reconstructed data. The correlation integral $CI$ for the $d$ dimensional trajectory is then evaluated, which measures the mean probability for different outputs to be within a distance $R$\cite{Procaccia83}. The growth of $CI$ with $R$ is evaluated, an example of which is shown in Figure S10, and the power-law exponent is determined. The exponent is called the correlation dimension $\nu$ of the system and saturates to a fixed value determined by the degrees of freedom of the system as $d$ is increased. For stochastic processes, $\nu$ does not exhibit any saturation and keeps increasing as $d$ is increased\cite{Procaccia83}.

\subsection{NIST SP 800-90B tests}
In order to run the SP 800-90B tests, a sequential dataset consisting of $10^6$ samples and a restart dataset generated from $10^3$ restarts of the device with $10^3$ samples from each restart is required. Restart dataset consists of two $10^3\times 10^3$ matrices where the different restarts correspond to the different rows or columns of the matrices. The testing phase starts with assessing the IID assumption against the data. A set of chi-square tests and permutation tests are executed to check for the independence and uniformity in the distribution of the data. If the sequential, restart rows, and restart columns datasets pass all the tests, the random sequence generated by the device is certified to be IID. If any of the tests are failed, the output is assumed to be non-IID.

The next stage in the validation process is estimating the min-entropy of the TRNG. Different methods are employed for calculating min-entropy dependent on the results from the IID test. For IID sequences, min-entropy is calculated using a most common value (MCV) estimator\cite{Turan18}. For non-IID sequences, the MCV estimator tends to overestimate the entropy and requires the use of multiple estimators for min-entropy assessment. To calculate the MCV entropy, the proportion of the most frequent value $\hat{P}$ is estimated as
\begin{equation}
\hat{P}=\max\limits_{i}\left\lbrace P(x_i)\right\rbrace,~~~~i=1,2,...,k
\end{equation}
An upper bound on the probability $P_u$ is estimated with a critical $Z_{1-0.005}$ value as
\begin{equation}
P_u=\min\left\lbrace 1,\hat{P}+2.576\sqrt{\frac{\hat{P}(1-\hat{P})}{Number~of~samples-1}}\right\rbrace
\end{equation}
The MCV entropy is then estimated as $H_{MCV}=-log_2P_u$. $H_{MCV}$ is calculated both for the sample output stream ($H_{symbol}$) and the binary stream ($H_{bitstring}$) and the minimum of $\left(H_{symbol}, n\times H_{bitstring}\right)$ is assigned as the min-entropy of the TRNG for the particular dataset. The next step is to validate the true random nature of the output against multiple restarts of the system. The restart tests consist of a sanity check and an entropy estimation\cite{Turan18}. The sanity check tests for any deviation in the entropy from the min-entropy obtained in the previous stages after the restart of the system. If the device passes the sanity check, an entropy estimate is awarded based on a suitable entropy estimator, which for IID outputs is the MCV estimator. This process is done for both row and column datasets. The lowest of all the min-entropy obtained in the different stages of validation is assigned as the min-entropy of the TRNG.

\section{Supporting Information}
The Supporting Information is available free of charge at \textbf{link}.\\
Energy band diagram; Band alignment estimation; Gate leakage characterization; Modulation of distribution of bits; Shannon entropy modulation; Randomness analysis at $295~K$; Recurrence plot; Difference plot; Lag plot; Correlation dimension estimation; Health tests; NIST SP 800-22 results; Performance benchmarking. (\textbf{link})

\section{Acknowledgements}
This work was supported in part by a Core Research Grant from the Science and Engineering Research Board (SERB) under Department of Science and Technology (DST), a grant from Indian Space Research Organization (ISRO), a grant from MHRD under STARS, and a grant from MHRD, MeitY and DST Nano Mission through NNetRA. K.W. and T.T. acknowledge support from the Elemental Strategy Initiative conducted by the MEXT, Japan (Grant Number JPMXP0112101001) and  JSPS KAKENHI (Grant Numbers 19H05790, 20H00354 and 21H05233).

\section{Competing Interest}
The authors declare no financial or non-financial competing interest.

\bibliography{ref2}

\newpage
\newpage
\newpage
\newpage
\AtEndDocument{

\section*{Supplementary material}

# Supplementary Figures

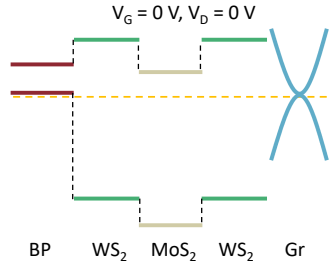

Figure S1: **Energy band diagram.** Band alignment of the heterostructure at $V_G = 0\ V$.

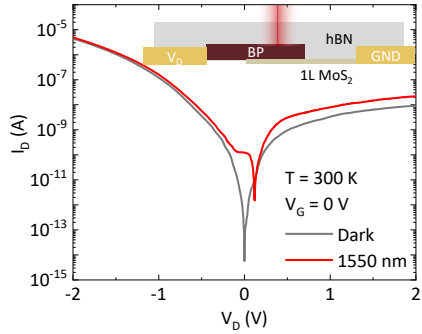

Figure S2: **Band alignment estimation.** $I_D - V_D$ characterization of a BP (drain)/1L-MoS$_2$ (source) stack under dark (gray trace) and 1550 nm laser excitation (red trace) exhibiting a positive $V_{OC}$. Inset: Schematic representation of the device.

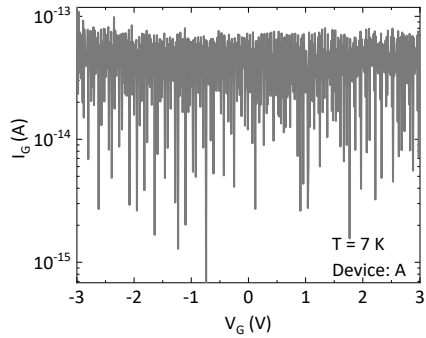

Figure S3: **Gate leakage characterization.** Observed gate leakage through the device showing a sub 0.1 pA current.

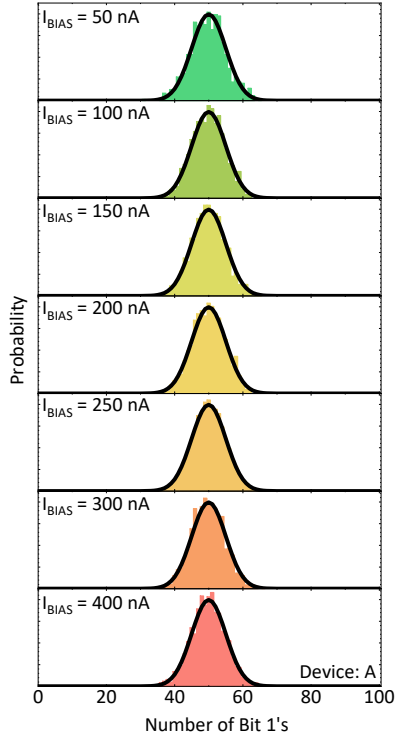

Figure S4: **Modulation of distribution of bits.** Measured distribution of the number of *bit* 1*s* in a block of 100 *bits* wide as a function of $I_{BIAS}$. The solid black line shows a theoretical binomial distribution with equal probability for *bit* 1 and *bit* 0, and the colored bars represent the experimental data.

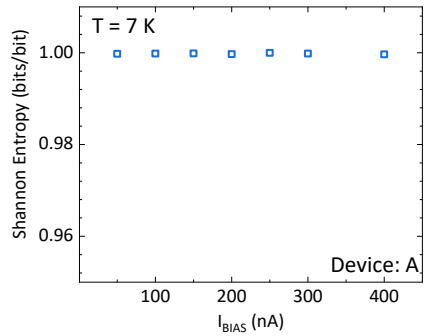

Figure S5: **Shannon entropy modulation.** Extracted Shannon entropy modulation as a function of $I_{BIAS}$ showing virtually no degradation of entropy with increasing $I_{BIAS}$.

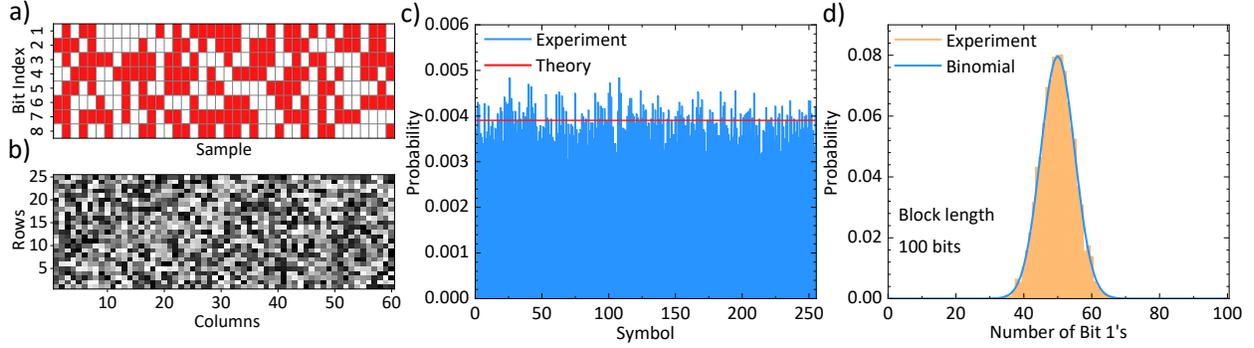

Figure S6: **Randomness analysis at** $295\ K$. **a)** A slice of the output stream obtained after digitizing the time bins with 8 bit resolution. White (red) squares represent bit 1 (bit 0). **b)** Grayscale map obtained by reshaping a stream of 1500 samples into a $25 \times 60$ matrix. 8 bit values ranging from 0 to 255 are mapped to the intensity of the pixel, with 255 corresponding to white. **c)** Probability distribution for all the symbols in the alphabet obtained experimentally (blue bars) and that of a theoretical uniform distribution (red trace). **d)** Probability distribution of the number of bit 1's in a binary block of length 100 bits. Experimental (orange bars) closely follow a theoretical unbiased binomial distribution (blue trace).

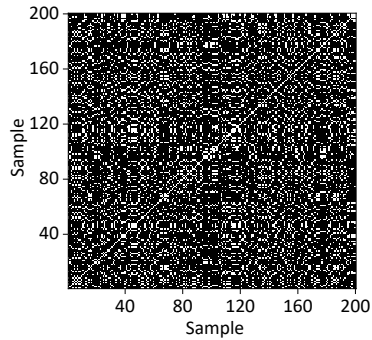

Figure S7: **Recurrence plot.** A sample recurrence plot obtained from 200 data points using a distance threshold of 10% of the maximum distance.

4

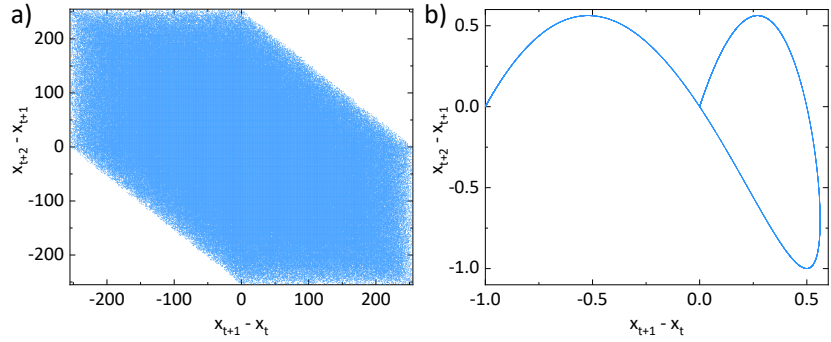

Figure S8: **Difference plot. a)** Difference plot obtained from the data showing no obvious pattern and filling all the mathematically allowed regions in the graph. **b)** Difference plot for a chaotic logistic map.

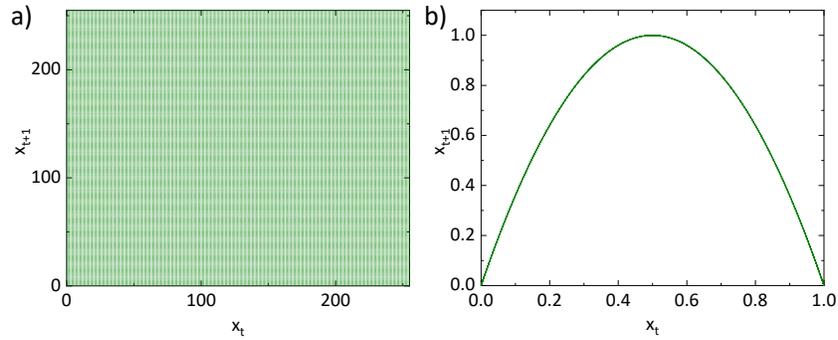

Figure S9: **Lag plot. a)** Lag plot showing the absence of any identifiable structures indicating the randomness of the data. **b)** Lag plot for a chaotic logistic map.

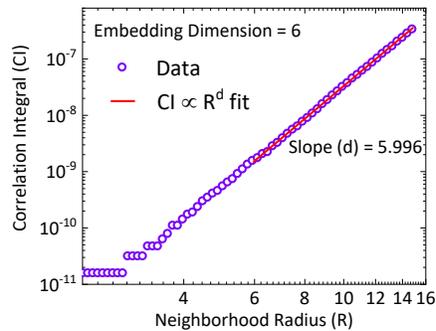

Figure S10: **Correlation dimension estimation.** Growth of the correlation integral of the data (violet circle markers) as a function of the neighborhood radius for an embedding dimension of 6 and a power law fit (red trace).

5

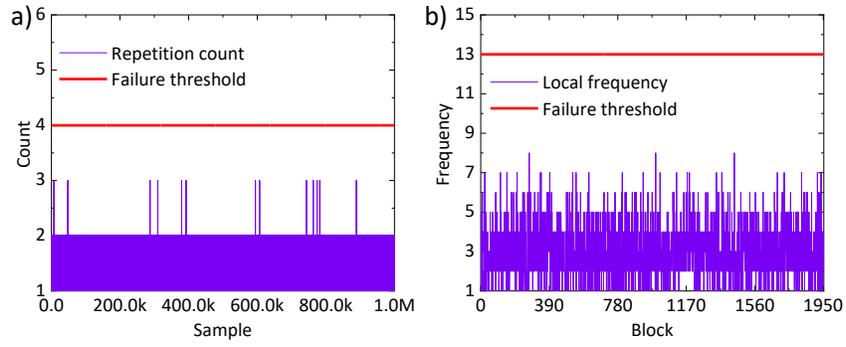

Figure S11: **Health tests. a)** Repetition count for the generated output stream (violet) and the failure threshold (red) for a false positivity rate $\alpha = 2^{-20}$. **b)** Frequency of randomly chosen symbol in a 512 sample wide block (violet) and the failure threshold (red) for a false positivity rate $\alpha = 2^{-20}$.

# Supplementary Tables

Table S1: **NIST SP 800-22 results**. Pass rate indicates the fraction of sequences tested which passed the particular test. The results field indicate whether the obtained pass rate lies in the acceptable range for the recommended significance level. For some tests, the number of sequences tested is lower than 100 due to the large data requirements of the particular test. Together with the pass rate analysis, a $\chi^2$ goodness of fit p-value is also typically obtained from the test suit, the aim of which is to analyze the uniformity of the data. For our TRNG, as the uniformity is confirmed by more exhaustive tests in NIST SP 800-90B, the p-value analysis of NIST SP 800-22 is not done.

| Test | Pass rate | Result |
| --- | --- | --- |
| The Frequency (Monobit) Test | 100/100 | Pass |
| Frequency Test within a Block | 100/100 | Pass |
| The Runs Test | 99/100 | Pass |
| Tests for the Longest-Run-of-Ones in a Block | 100/100 | Pass |
| The Binary Matrix Rank Test | 100/100 | Pass |
| The Discrete Fourier Transform (Spectral) Test | 99/100 | Pass |
| The Non-overlapping Template Matching Test | Pass | Pass |
| The Overlapping Template Matching Test | 8/8 | Pass |
| Maurer's "Universal Statistical" Test | 20/20 | Pass |
| The Linear Complexity Test | 8/8 | Pass |
| The Serial Test | Pass | Pass |
| The Approximate Entropy Test | 98/100 | Pass |
| The Cumulative Sums (Cusums) Test | 100/100 | Pass |
| The Random Excursions Test | 4/4 | Pass |
| The Random Excursions Variant Test | 4/4 | Pass |

Table S2: **Performance benchmarking.** The device platform, the IID nature, results from the SP 800-22 and the SP 800-90B tests and the extracted min-entropy/ bit are compared wherever the data is available. The IID nature of the data is confirmed from the results of SP 800-9B tests. For the cases without data from SP 800-90B, the IID nature is evaluated based on results from SP 800-22.

| Ref | Platform | IID | SP 800-22 | SP 800-90B | Min-Entropy (bits/bit) |
|---|---|---|---|---|---|
| (3) | Optical | No | | Pass | 0.22 |
| (4) | Electronic | No | Pass | Pass | 0.893 |
| (5) | Optical | Yes | Yes | | |
| (6) | Electronic | Yes | Yes | | |
| (9) | Optical | Yes | Yes | | |
| (11) | Super-paramagnetic | Yes | Yes | | |
| (23) | Electronic | Yes | Yes | | |
| (30) | Electronic | No | Pass | Pass | 0.889 |
| (31) | Electronic | No | Pass | Pass | 0.721 |
| (32) | Optical | No | | Pass | 0.219 |
| (33) | Electronic | Yes | Yes | | |
| (34) | Electronic | Yes | Yes | | |
| **This Work** | **Electronic** | **Yes** | **Pass** | **Pass** | **0.983** |

8

}
\end{document}